\documentclass[a4paper,11pt]{article}
\usepackage{amsmath,amsfonts,amssymb,times,mathptmx,graphicx,amsthm,lineno}
\usepackage{placeins}
\usepackage[T1]{fontenc}
\usepackage[latin1]{inputenc}
\title{\bf
The Cramer-Rao inequality to go beyond the $\mathbf{\sqrt{N}}$-limit of the
standard least-squares method in track fitting.
 }
\author{Gregorio Landi$^a$\thanks{Corresponding
author. Gregorio.Landi@fi.infn.it}~,   Giovanni E. Landi$^b$\\
\\
\llap{$^a$} Dipartimento di Fisica e Astronomia,
Universita' di Firenze and INFN\\
Largo E. Fermi 2 (Arcetri) 50125, Firenze, Italy\\
\\
\llap{$^b$} ArchonVR S.a.g.l.,\\
Via Cisieri 3,
6900 Lugano, Switzerland.\\ \\
{ October 30, 2019}}
\date{ }
\begin{document}
\maketitle 
\begin{abstract}
The Cramer-Rao-Frechet inequality is reviewed specializing it to
track fitting.  
A diffused opinion 
attributes to this inequality the 
limitation of the resolution of the track fits  with the number N 
of observations. It turns out that this opinion is incorrect,  weighted least
squares method is not subjected to that N-limitation. 
In a previous publication, simulations
with a simple Gaussian model produced
interesting results: a linear growth of the peaks 
of the distributions with the number $N$ of observations, 
much faster than the $\sqrt{N}$ of the standard least squares.
These results could be considered  a violation of a well known 
$1/N$-rule for the variance of an unbiased estimator, frequently reported 
as the Cramer-Rao-Frechet bound.
To clarify this point beyond any doubt, it would be essential a direct
proof of the consistency of those results with this 
inequality. Unfortunately, such proof is lacking or very difficult to find.
Hence, the Cramer-Rao-Frechet developments are
applied to prove the efficiency (optimality) of the simple 
Gaussian model and the consistency of its results. The inequality remains valid even for irregular
models supporting the results of realistic models with similar growths. 

\end{abstract}



\tableofcontents

\pagenumbering{arabic} \oddsidemargin 0cm  \evensidemargin 0cm


\section{Introduction}\indent

Very recently criticisms and doubts were raised against our
paper (ref.~\cite{landi07}) on track fitting.
The contested results were principally those of the following fig.~\ref{fig:figure_1}.
This figure illustrated (Monte Carlo) simulations of a very simple tracker with N
detecting layers of identical technology (silicon microstrips) crossed by a set of
parallel straight tracks of minimum ionizing particles (MIPs). The reported
estimator was the track direction.
\begin{figure} [h!]
\begin{center}
\includegraphics[scale=0.75]{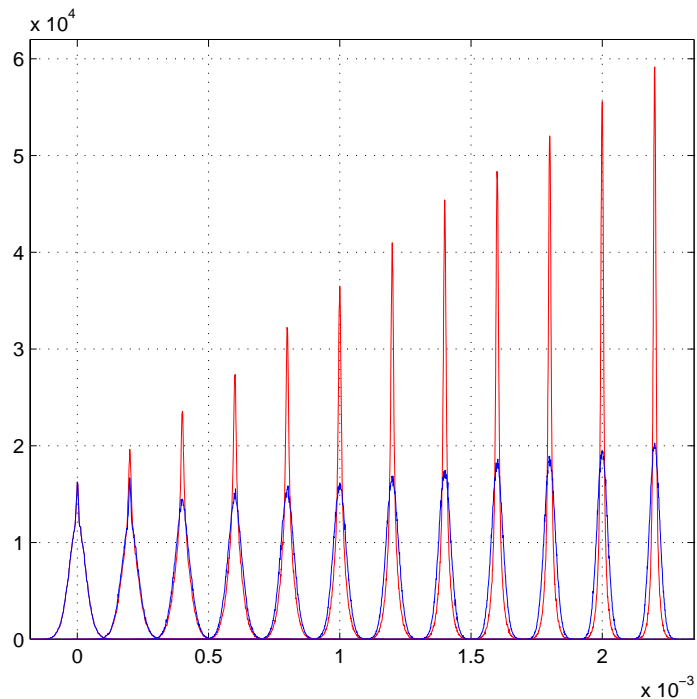}
\caption{\em The simple Gaussian model of ref.[1]. Distributions of the fits to the
track directions for tracker models with N=2 to N=13 detecting layers.
The first distributions are centered on zero, the others are shifted of N-2 steps of fixed amplitude.
The blue distributions are the results of the standard least-squares. The red distributions are 
obtained with the weighted least-squares. }
\label{fig:figure_1}
\end{center}
\end{figure}

The hit uncertainties were approximated with Gaussian distributions with two type
of hit quality (heteroscedastic model). One of very good quality when the MIP
crosses the detector near to the borders ($20\%$ of strip width), and the other
of bad quality ($80\%$ of the strip width) around the strip center. A slight
random mechanical misalignment was supposed with mean value around a half strip (30 $\mu m$).
Random misalignments much larger than this are always present in real trackers.
This misalignment (rotations and translations) is simulated as
a randomization of the hit position on the strips producing a binomial distribution of
the hit quality. The collection of N hits (a track) is fitted with two different
least squares method, one with the standard least squares and the other with the
weighted least-squares with the weights given by the hit quality.  The track direction
is the illustrated estimator.
With  standard least-squares, we define the usual least-squares where all the used
observations have identical variances (homoscedastic models). The standard least-squares,
built for homoscedastic models, when applied to heteroscedastic models gives results
almost insensitive to the presence of the good hits, and the maximums of the empirical
probability density functions (PDFs) show a growth very near to $\sqrt{N}$. The blue
lines of fig.1. Instead, weighted least-squares, consistent for the heteroscedastic
models, reports a surprising fast increase, almost linear, of the maximums of the
simulated PDFs. Translating these results on mean variances, the standard fit
shows a $1/N$ trend and the weighted fit shows a trend of the mean variances as $1/N^2$.
This trend $1/N^2$  was considered a violation of the Cramer-Rao-Frechet (CRF) bound
as reported in many books and documentations. Our arguments about the different hypotheses
used to prove the $1/N$ rule and those of our approach had no effect.
To settle any type of contrast,  a complete
demonstration is required to prove our consistency with the CRF-inequality.
It is evident that this long demonstration could not be inserted in ref.~\cite{landi07}.
The main part of ref.~\cite{landi07} deals with physical models with Cauchy-like
tails and (in theory) with infinite variance. Hence, any finite bound on the variance
given by the CRF-inequality is automatically satisfied by them. However, the strong
similarity of the physical models with the Gaussian model could extend them the suspects
of possible inconsistencies.  Thus, to give a strong support to our results we review
the CRF inequality to show the inappropriate extension of the rule $1/N$ to our approach.
For this, we will apply the CRF developments to heteroscedastic models and to the least
squares method for track fitting (for straight tracks and curved tracks) with a special 
attention to our Gaussian model. The last two points are very difficult to find
in literature. Surely other authors did similar developments
and we apologize to not citing them. In any case, the enormous statistical samples of
tracks (of the order of $10^{12}$ tracks per year) renders very important  careful studies
of any detail of their fitting methods. During our analysis of the track fitting we had
to read carefully many papers (in any case a small portion of those published) on this
subject and we found inaccurate fitting strategies. Probably the reason of this
inaccuracy is connected to the weak growth as $\sqrt{N}$ of the standard least-squares
(the Kalmann filter is essentially identical) that is negligibly modified by any
improvement. Another element of inaccuracy is a frequent claim of a "measure" the hit resolutions.
In reality, those parameters are not a "measure", but an application of 
an equation valid only for homoscedastic models, inappropriate 
for the physics of the trackers (surely always heteroscedastic).
Instead, our fitting methods of ref.~\cite{landi07,landi06,landi05} open very different 
possibilities.
The heteroscedastic models allow drastic improvements of the quality of the track
fitting. In spite of the possible doubts, they are perfectly consistent with the
CRF inequality. Instead, the CRF inequality demonstrates
the inefficiency of the standard least squares for heteroscedastic models. The lucky
model of ref.~\cite{landi07} is a first order tool to insert heteroscedasticity 
in track fitting and able to produce consistent improvements.

\section{The standard CRF-inequality}

The developments of this section follow as far as possible ref.~\cite{mathstat} and ref.~\cite{gowan}.
As usual it is considered a N-times repetition of random observations $\{x_1,x_2\ldots x_N\}$ described by an identical
probability density function (PDF) $f(x,\theta)$, where $\theta$ is a (scalar) parameter. These
random variables $\mathbf{x}=\{x_1,x_2\ldots x_N\}$ are  indicated as independent and identically distributed or
i.i.d. random variables. We will use this notation in the following, without commenting the
absolute physical inconsistency of this assumption. The probability of the set $\mathbf{x}$ of i.i.d.
random variables (likelihood function) is:
\begin{equation}\label{eq:equation_1}
    L(\mathbf{x},\theta)=f(x_1,\theta)f(x_2,\theta)\ldots\ldots f(x_N,\theta)\,.
\end{equation}
As usual $f(x,\theta)$ is positive, normalized and supposed differentiable with respect 
to the $\theta$-parameter.
Let us define the random variable $U(\mathbf{x},\theta)$:
\begin{equation}
    U(\mathbf{x},\theta)=\frac{\partial \ln\, L(\mathbf{x},\theta)}{\partial\theta}=
    \sum_{i=1}^N\frac{\partial \ln\, f({x_i},\theta)}{\partial\theta}\,.
\end{equation}
The following developments require to differentiate the integrals of $f({x},\theta)$
and invert the order of integration and differentiation. The functions that allow these operations
are called regular models and they imply strict analytical conditions. A necessary condition of
regular models is the independence of the range of $f({x},\theta)$ from the $\theta$-parameter.
The normalization of the $f({x},\theta)$ is:
\begin{equation}\label{eq:equation_4}
    \int  L(\mathbf{x},\theta) \mathrm{d}\mathbf{x}=
    \int f(x_1,\theta)f(x_2,\theta)\ldots\ldots f(x_N,\theta)d x_1 d x_2 \ldots d x_N=1\,.
\end{equation}
Differentiating eq.~\ref{eq:equation_4} with respect to $\theta$ and inverting the integral
with the differentiation gives:

\begin{equation}\label{eq:equation_5}
\begin{aligned}
   0=&\int \frac{\partial L(\mathbf{x},\theta)}{\partial \theta} \mathrm{d}\mathbf{x}=
   \sum_{i=1}^N\int \frac{\partial f({x_i},\theta)}{\partial \theta} \mathrm{d}{x_i}=\\
   &\int \frac{\partial \ln L(\mathbf{x},\theta)}{\partial \theta}L(\mathbf{x},\theta) \mathrm{d}\mathbf{x}=
   \sum_{i=1}^N\int \frac{\partial \ln f({x_i},\theta)}{\partial \theta}f({x_i},\theta) \mathrm{d}{x_i}\,.\\
\end{aligned}
\end{equation}
The second line of eq.~\ref{eq:equation_5} is given by a property of the logarithmic derivative and
it implies that the mean value of $U(\mathbf{x},\theta)$ is zero.
Instead, its mean square is different from zero:
\begin{equation}\label{eq:equation_6}
    \int U(\mathbf{x},\theta)^2 \mathrm{d} \mathbf{x}=
    \sum_{i=1}^N\int \big(\frac{\partial \ln f({x_i},\theta)}{\partial \theta}\big)^2f({x_i},\theta) \mathrm{d}{x_i}=
    N \,\int \big(\frac{\partial \ln f({x},\theta)}{\partial \theta}\big)^2f({x},\theta) \mathrm{d}{x}\,.
\end{equation}
The first term of eq.~\ref{eq:equation_6} is called Fisher information of the sample, the last integral is the
Fisher information contained in a single observation.\noindent

{\bf
For the i.i.d. random variables the Fisher information is
N-times the Fisher information of a single observation. }

This originates  the $1/N$ factor in the CRF-bound. It will evident that for
non i.i.d (heteroscedastic) random variables this factorization disappears.
We underline the i.i.d. assumption in eq.~\ref{eq:equation_6} because
often this condition is a standard assumption (as in ref.~\cite{mathstat}), and it can only be recovered
with an accurate re-reading of the intere demonstration and doing another
demonstration without the i.i.d assumption.

If $f(x,\theta)$ is twice differentiable with respect to $\theta$, the last term of eq.~\ref{eq:equation_5}
gives the following identity:
\begin{equation}\label{eq:equation_7}
    0=\int \big(\frac{\partial \ln f({x},\theta)}{\partial \theta}\big)^2f({x},\theta) \mathrm{d}{x}+
    \int \frac{\partial^2 \ln f({x},\theta)}{\partial \theta^2}\,f({x},\theta) \mathrm{d}{x}\,.
\end{equation}
The Fisher information $i(\theta)$ for a single observation becomes:
\begin{equation}
    i(\theta)=\int \big(\frac{\partial \ln f({x},\theta)}{\partial \theta}\big)^2f({x},\theta) \mathrm{d}{x}=
    -\int \frac{\partial^2 \ln f({x},\theta)}{\partial \theta^2}\,f({x},\theta) \mathrm{d}{x}\,.
\end{equation}
The last form of $i(\theta)$ is particularly effective for Gaussian PDF.

\subsection{The CRF inequality}

At this point is is easy to obtain the CRF-inequality for these i.i.d. random variables.
Defining an estimator $T(\mathbf{X})=T(x_1,x_2\ldots x_n)$ and its mean value:
\begin{equation}\label{eq:equation_9}
    \int T(\mathbf{x}) L(\mathbf{x},\theta) \mathrm{d} \mathbf{x}=\tau(\theta)\,,
\end{equation}
differentiating eq.~\ref{eq:equation_9} in $\theta$ we obtain:
\begin{equation}
\begin{aligned}
    &\tau(\theta)'=\int T(\mathbf{x})\frac{\partial \ln L(\mathbf{x},\theta)}{\partial \theta}L(\mathbf{x},\theta) \mathrm{d}\mathbf{x}=\int \big[T(\mathbf{x})-\tau(\theta)\big]\frac{\partial \ln L(\mathbf{x},\theta)}{\partial \theta}L(\mathbf{x},\theta) \mathrm{d}\mathbf{x}=\\
    &\int \widehat{T}(\mathbf{x})\sqrt{L(\mathbf{x},\theta)}
    \frac{\partial \ln L(\mathbf{x},\theta)}{\partial \theta}
    \sqrt{L(\mathbf{x},\theta)} \mathrm{d}\mathbf{x} \ \ \ \ \ \ \widehat{T}(\mathbf{x})={T}(\mathbf{x})-\tau(\theta)\,.\\
\end{aligned}
\end{equation}
The form of the last integral is allowed by the positivity of $L(\mathbf{x},\theta)$.
The subtraction of $\tau(\theta)$ from $T(\mathbf{x})$ does not modify the integral
for eq.~\ref{eq:equation_5}.   The Cauchy-Schwarz inequality
gives:
\begin{equation}\label{eq:equation_11}
\begin{aligned}
    &[\tau(\theta)']^2\leq \int \widehat{T(\mathbf{x})}^2{L(\mathbf{x},\theta)} \mathrm{d}\mathbf{x}\int\big[\frac{\partial \ln L(\mathbf{x},\theta)}{\partial \theta}\big]^2L(\mathbf{x},\theta) \mathrm{d}\mathbf{x}\\
    &[\tau(\theta)']^2\leq\,N\, i(\theta) \int \widehat{T(\mathbf{x})}^2{L(\mathbf{x},\theta)} \mathrm{d}\mathbf{x}
\end{aligned}
\end{equation}
or in its final form:
\begin{equation}\label{eq:equation_12}
    \int \widehat{T(\mathbf{x})}^2{L(\mathbf{x},\theta)} \mathrm{d}\mathbf{x}\geq
    \frac{[\tau(\theta)']^2}{N\, i(\theta)}\,.
\end{equation}
The variance of $T(\mathbf{x})$ can not be lower than the right side of eq.~\ref{eq:equation_12}.
This is the well-known CRF inequality that creates many doubts
to our readers. When the CRF inequality becomes an identity, the estimator is defined efficient.

\subsection{Efficient estimators}

The Cauchy-Schwarz inequality becomes an equality if and only if $\widehat{T}(\mathbf{x})$ and
$U(\mathbf{x},\theta)$ are linearly dependent:
\begin{equation}\label{eq:equation_13}
    T(\mathbf{x},\theta)-\tau(\theta)=a(\theta)U(\mathbf{x},\theta)
\end{equation}
Hence, if an efficient estimator exists, it is a function of the model. The
logarithmic functions contained in $U(\mathbf{x},\theta)$ allow to find efficient estimators
only in exponential models.

\subsection{The standard mean with an i.i.d. Gaussian model}

 The PDF $f(x,\theta)$, the mean value $\tau(\theta)$ and the Fisher information $i(\theta)$ are:

\begin{equation}\label{eq:equation_14}
\begin{aligned}
    &f(x,\theta)=\frac{\exp[-(x-\theta)^2/2\sigma^2]}{\sqrt{2\pi}\sigma}\ \ \ \ \ \ \ \tau(\theta)=\int T(\mathbf{x})L(\mathbf{x},\theta) \mathrm{d} \mathbf{x}=\theta\\
    &i(\theta)=-\int \frac{\partial^2 \ln f({x},\theta)}{\partial \theta^2}\,f({x},\theta) \mathrm{d}{x}=\frac{1}{\sigma^2}\\
    &U(\mathbf{x},\theta)=\frac{1}{\sigma^2}\sum_{j=1}^N\,x_j-\frac{N}{\sigma^2}\theta \ \ \ \ \ \ \frac{1}{N i}U(\mathbf{x},\theta)=
    T^e(\mathbf{x})-\theta \ \ \ \ \ T^e(\mathbf{x})=\frac{\sum_{j=1}^N\,x_j}{N}
\end{aligned}
\end{equation}
The last of eq.~\ref{eq:equation_14} gives the efficient estimator of $\theta$ for this model as
defined by eq.~\ref{eq:equation_13}. Its unbiased form $\widehat{T^e(\mathbf{x})}=T^e(\mathbf{x})-\theta$,
inserted in eq.~\ref{eq:equation_12} gives the identity $\sigma^2/N=\sigma^2/N$.

\section{CRF inequality for heteroscedastic models}

It is well-known by long time that different observations have always non identical PDF.
The universe would be very simple if the observations were i.i.d., in this case the
precision of a measure could be improved at any level by the the simple repetition and averaging the
results. Gauss knew very well this problem and in his paper of 1823 extended his
least squares method to handle observations of different PDF.
For our misfortune the standard books of mathematical statistics do not
cover this argument. But to complete our confutation of critics moved
to ref.~\cite{landi07}, we have to venture in this field.
As in the previous section,
we suppose a set of observations obtained by an array of different detectors as in a tracker.
Even if all the detectors are built with identical technology, differences are always present.
Each detector is anisotropic for its construction, optimized to the measure of positions.
The signal spreads in a different recognizable way for each hit point.
Moreover, each
observation has a quantum mechanical interaction that renders unrepeatable the observations.
However, a PDF for each observation can be defined (ref.~\cite{landi05}) even if very
different from the form required for the CRF-inequality. However, the likelihood of a set
of heteroscedastic observations becomes:
\begin{equation}\label{eq:equation_15}
    L_{1,2,\ldots,N}(\mathbf{x},\theta)=f_1(x_1,\theta)f_2(x_2,\theta)\ldots\ldots f_N(x_N,\theta)
\end{equation}
The evident difference is an index attached to each PDF. Instead of the
awful PDFs of ref.~\cite{landi05,landi06}, we will use an easy model.
The simple model of ref.~\cite{landi07}
is a heteroscedastic Gaussian model (independent non identically distributed
Gaussian random variables), hence, it has all the analytical properties required for the
CRF-inequality. Again $U_{1,2,\ldots,N}(\mathbf{x},\theta)$ is defined as:
\begin{equation}\label{eq:equation_16}
    U_{1,2,\ldots,N}(\mathbf{x},\theta)=\frac{\partial \ln\, L_{1,2,\ldots,N}(\mathbf{x},\theta)}{\partial\theta}=
    \sum_{i=1}^N\frac{\partial \ln\, f_i({x_i},\theta)}{\partial\theta}
\end{equation}
Even the Fisher information acquires a set of indices,
\begin{equation}\label{eq:equation_17}
    \int U_{1,2,\ldots,N}(\mathbf{x},\theta)^2 L_{1,2,\ldots,N}(\mathbf{x},\theta)\mathrm{d} \mathbf{x}=
    \sum_{j=1}^N\int \big(\frac{\partial \ln f_j({x_j},\theta)}{\partial \theta}\big)^2f_j({x_j},\theta) \mathrm{d}{x_j}=
    \sum_{j=1}^N i_j(\theta)
\end{equation}
but now all the PDF of the observations are different as the Fisher information of each
observation. The factorization of eq.~\ref{eq:equation_6} disappears with its
N factor and its supposed generality.
All the other equations are formally identical up to the first line of 
eq.\ref{eq:equation_11} ( a part an additional index to
the PDFs ).
With the new Fisher information the CRF-inequality for an unbiased estimator
$\widehat{T}_{1,2,\ldots,N}(\mathbf{x})$ becomes:
\begin{equation}
    \int \widehat{T}_{1,2,\ldots,N}(\mathbf{x})^2{L_{1,2,\ldots,N}(\mathbf{x},\theta)} \mathrm{d}\mathbf{x}\geq
    \frac{[\tau_{1,2,\ldots,N}(\theta)']^2}{\sum_{j=1}^N i_j(\theta)}
\end{equation}
Supposing a large set of possible PDFs ( as those of ref.~\cite{landi05} ) the
estimator $T_{1,2,\ldots,N}(\mathbf{x},\theta)$ must keep the track of
the types and the order of the PDFs contained. It must be
substituted with $T_{j_1,j_2,\ldots j_N}(\mathbf{x},\theta)$, similarly for the likelihood
function $L_{j_1,j_2,\ldots j_N}(\mathbf{x},\theta)$ and the Fisher information.
In this extended form the CRF- inequality becomes:
\begin{equation}
    \int \widehat{T}_{j_1,j_2,\ldots j_N}(\mathbf{x})^2{L_{j_1,j_2,\ldots j_N}(\mathbf{x},\theta)} \mathrm{d}\mathbf{x}\geq
    \frac{[\tau_{j_1,j_2,\ldots j_N}(\theta)']^2}{i_{j_1,j_2,\ldots j_N}(\theta)}
\end{equation}
The efficient estimators have the variance identical to the CRF-bound. Again the logarithmic
functions of eq.~\ref{eq:equation_16} limit the definition of efficient estimators to exponential PDFs.

\subsection{The weighted mean in heteroscedastic Gaussian models}

Let us consider the standard mean $T_{1,2,\ldots N}^a(\mathbf{x})=\sum_{i=1}^N x_i/{N}$
as an estimator in heteroscedastic Gaussian models.
The PDFs $f_j(x,\theta)$, the mean value $\tau_{1,2,\ldots N}(\theta)$ and
the Fisher information due to the j-PDF   $i_j(\theta)$ are:
\begin{equation}\label{eq:equation_20}
\begin{aligned}
    &f_j(x,\theta)=\frac{\exp[-(x-\theta)^2/2\sigma_j^2]}{\sqrt{2\pi}\sigma_j}\ \ \ \ \ \ \ \tau_{1,2,\ldots N}(\theta)=\int T_{1,2,\ldots N}^a(\mathbf{x})L_{1,2,\ldots N}(\mathbf{x},\theta) \mathrm{d} \mathbf{x}=\theta\\
    &i_j(\theta)=-\int \frac{\partial^2 \ln f_j({x},\theta)}{\partial \theta^2}\,f_j({x},\theta)
    \mathrm{d}{x}=\frac{1}{\sigma_j^2} \ \ \ \ \ \ \int\big[\sum_{i=1}^N \frac{x_i}{N}-\theta\big]^2
    \,L_{1,2,\ldots N}(\mathbf{x},\theta) \, \mathrm{d} \mathbf{x} =\frac{\sum_{J=1}^N\sigma_j^2}{N^2}
\end{aligned}
\end{equation}

With the variance of $T^a$ the CRF-inequality is:
\begin{equation}
    \frac{\sum_{J=1}^N\sigma_j^2}{N^2}>\frac{1}{\sum_{j=1}^N\frac{1}{\sigma_j^2}}
\end{equation}
The equality is impossible for the assumption of the heteroscedasticity (the identity of the $\sigma_j$ is excluded).
Let us extract from eq.~\ref{eq:equation_16} an efficient estimator defined (if it exists) as usual:
\begin{equation}
    T_{1,2,\ldots N}(\mathbf{x})-\tau_{1,2,\ldots N}(\theta)=a(\theta) U_{1,2,\ldots,N}(\mathbf{x},\theta) \ \ \ \  \ \ \
\end{equation}
With $a(\theta)=1/[\sum_{j=1}^N 1/\sigma_j^2]$ and $\tau_{1,2,\ldots N}=\theta$ it is:
\begin{equation}
     \ \  \
    a(\theta)\sum_{i=1}^N\frac{\partial \ln\, f_i({x_i},\theta)}{\partial\theta}=\frac{1}{\sum_{j=1}^N 1/\sigma_j^2}\sum_{i=1}^N\frac{x_i}{\sigma_i^2}-\theta
    \ \ \ \ \Rightarrow \ \ T_{1,2,\ldots N}^e(\mathbf{x})=
    \sum_{i=1}^N\frac{x_i}{\sigma_i^2}\frac{1}{\sum_{j=1}^N\frac{1}{\sigma_j^2}}
\end{equation}
This estimator is proportional to $\partial\ln L_{1,2,\ldots,N}(\mathbf{x},\theta)/\partial \theta$
and transforms the Cauchy-Schwarz inequality in an identity.
It is easy to verify, without the CRF inequality, that the variance of
the standard mean is always greater than the variance of the weighted mean.
\begin{equation}
  \frac{ \sum_{J=1}^N\sigma_j^2}{N^2}>\frac{1}{\sum_{j=1}^N\frac{1}{\sigma_j^2}}  \ \ \ \ \ \ \ \ \ \ \
  \ \sum_{J=1}^N\sigma_j^2\sum_{j=1}^N\frac{1}{\sigma_j^2}>N^2
\end{equation}
The second equation is the Cauchy-Schwarz inequality for the two vectors 
$\{\sigma_j\}$ and $\{1/\sigma_j\}$ . The equality
is impossible because $\sigma^2$ and $1/\sigma^2$ can never be proportional excluding the trivial
case of identical $\sigma_j$.  Heteroscedasticity excludes this case.
Thus, in heteroscedastic models, the weighted average has always minor variance
compared to the standard mean for regular and irregular models.

Hence, each combination of observations from our Gaussian model satisfies the
CRF inequality. No $1/N$ factor  is present in the
equation for a different Fisher information of each single observation. The case
of identity of all the observations is excluded {\em a priori}. Moreover, the mean
of the standard variances of a set of random sequences $\{j_1,j_2\ldots,j_N\}$, not
all identical, is higher than the mean of the corresponding weighted averages.
The absence of $1/N$ factors in the variance of the weighted average allows fast
decreases with $N$ that is impossible to the variance of the standard mean.

\subsection{The weighted mean in a homoscedastic model}

To prove another aspect of the CRF-inequality, we use the weighted mean outside its 
definition model (heteroscedasticity) and we use it in a homoscedastic model. In this case
the standard mean is the efficient estimator. The CRF-inequality states that the variance
of the weighted mean is greater than the variance of the standard mean. It is easy to 
illustrate this fact (the reverse of the last equations).
\begin{equation}
    \int \left (T_{1,2\ldots,N}^e(\mathbf{x})-\theta\right)^2 L(\mathbf{x},\theta)\, \mathrm{d} \mathbf{x}
    =\sigma^2\frac{\sum_j\,1/\sigma_j^4}{(\sum_l 1/\sigma_l^2)^2} > \frac{\sigma^2}{N}
    \ \ \ \ \ \ \Rightarrow\ \ \ \ \  N \sum_j \frac{1}{\sigma_j^4}\,>\, \Big(\sum_l\frac{1}{\sigma_l^2}\Big)^2
\end{equation}
The last inequality is the Cauchy-Schwarz inequality. The equality is excluded. 
Thus an efficient estimator in a model, outside its model has always a greater variance  
compared to the efficient estimator of the new model.

\section{CRF-bound for heteroscedastic least squares of straight tracks}

The application of the CRF-inequality to heteroscedastic least-squares fit for straight tracks  adds other
complications. Two parameters must be handled. The first parameter is the constant $\beta$:
the impact point of the track in the reference plane. The second is $\gamma\,y_i$: the angular shift of the
track in the plane distant $y_i$ from the reference plane. Also here, the heteroscedasticity eliminates
any easy rule with the increase of the number ($N$) of the observation planes
for two different reasons. One is again the
differences of the $\sigma_i$; the new one is due to the constraint of the fixed
length of the tracker. Each
insertion of a new detecting plane requires a repositioning of all the others. We will study our
Gaussian model with the condition $\sum_{j=1}^N y_j=0$, this introduces some simplification on
standard least squares equations. The likelihood function is similar to that
of eq.~\ref{eq:equation_15}, but now, the order of the PDFs must be conserved.
The parameter $y_j$ orders the measuring planes:
\begin{equation}\label{eq:equation_25}
    L_{1,2,\ldots,N}(\mathbf{x},\beta,\gamma)=f_1(x_1,\beta,y_1\gamma)f_2(x_2,\beta,y_2\gamma)\ldots\ldots f_N(x_N,\beta,y_N\gamma)
\end{equation}
\begin{equation}
\begin{aligned}
    &f_j(x,\beta,y_j\gamma)=\frac{\exp[-(x-\beta-y_j\gamma)^2/2\sigma_j^2]}{\sqrt{2\pi}\sigma_j}\ \ \ \ \ \ \ \tau_{1,2,\ldots N}^{\gamma}(\beta,\gamma)=\int T_{1,2,\ldots N}^{\gamma}(\mathbf{x})L_{1,2,\ldots N}(\mathbf{x},\beta,\gamma) \mathrm{d} \mathbf{x}=\gamma\\
    &\tau_{1,2,\ldots N}^{\beta}(\beta,\gamma)=\int T_{1,2,\ldots N}^{\beta}(\mathbf{x})L_{1,2,\ldots N}(\mathbf{x},\beta,\gamma) \mathrm{d} \mathbf{x}=\beta\\
\end{aligned}
\end{equation}
The function $U_{1,2,\ldots,N}(\mathbf{x},\theta)$ is the $2x1$ matrix $U_{1,2,\ldots,N}(\mathbf{x},\beta,\gamma)$:
\begin{equation}\label{eq:equation_27}
    U_{1,2,\ldots,N}(\mathbf{x},\beta,\gamma)=\left (\begin{array} {l}
    \displaystyle{\frac{\partial\ln L_{1,2,\ldots,N}(\mathbf{x},\beta,\gamma)}{\partial\beta}}\\
    \displaystyle{\frac{\partial\ln L_{1,2,\ldots,N}(\mathbf{x},\beta,\gamma)}{\partial\gamma}}
     \end{array} \right )
\end{equation}

The Fisher information $\mathbf{I}_{1,2,\ldots,N}$ becomes the $2x2$ matrix:
\begin{equation}\label{eq:equation_28}
\mathbf{I}_{1,2,\ldots,N}=\int \mathrm{d} \mathbf{x} L_{1,2,\ldots,N}(\mathbf{x},\beta,\gamma) \left (\begin{array} {l l}
     \displaystyle{-\frac{\partial^2\ln L_{1,2,\ldots,N}(\mathbf{x},\beta,\gamma)}{\partial\beta^2}}
    &  \displaystyle{-\frac{\partial^2\ln L_{1,2,\ldots,N}(\mathbf{x},\beta,\gamma)}{\partial\beta\partial\gamma}}\\
     \displaystyle{-\frac{\partial^2\ln L_{1,2,\ldots,N}(\mathbf{x},\beta,\gamma)}{\partial\gamma\partial\beta}}
    & \displaystyle{-\frac{\partial^2\ln L_{1,2,\ldots,N}(\mathbf{x},\beta,\gamma)}{\partial\gamma^2}}
\end{array}\right )
\end{equation}
\begin{equation}
    \mathbf{I}_{1,2,\ldots,N}=\left (\begin{array} {l l}
                     \textstyle{\sum_{j=1}^N\frac{1}{\sigma_j^2}} &  \textstyle{\sum_{j=1}^N\frac{y_j}{\sigma_j^2}}\\
                       &   \\
                      \textstyle{\sum_{j=1}^N\frac{y_j}{\sigma_j^2}} & \textstyle{\sum_{j=1}^N\frac{y_j^2}{\sigma_j^2}} \end{array} \right )
\end{equation}

Its inverse is:
\begin{equation}
    \mathbf{I^{-1}}_{1,2,\ldots,N}=\frac{1}{\mathrm{Det}\,\{\mathbf{I}_{1,2,\ldots,N}\}\,}\left (\begin{array} {l l}
                      \textstyle{\ \ \ \sum_{j=1}^N\frac{y_j^2}{\sigma_j^2}} &  \textstyle{-\sum_{j=1}^N\frac{y_j}{\sigma_j^2}}\\
                       \     &   \\
                      \textstyle{-\sum_{j=1}^N\frac{y_j}{\sigma_j^2}} &  \textstyle{\ \ \ \sum_{j=1}^N\frac{1}{\sigma_j^2}} \end{array} \right )
\end{equation}
The CRF inequality assumes the form of a matrix inequality, we limit to relations among variances:
\begin{equation}\label{eq:equation_31}
   Diag. \left (\begin{array} {l l l} Var(\beta,\beta) & & Cor(\beta, \gamma) \\
                                                & & \\
                                 Cor(\gamma,\beta) & & Var(\gamma,\gamma)\\
           \end{array} \right ) \geq Diag.\, \mathbf{I^{-1}}_{1,2,\ldots,N}
\end{equation}
where $Var(\beta,\beta)$ and $Var(\gamma,\gamma)$ are the variances of $\beta$ and $\gamma$:
\begin{equation}\label{eq:equation_32}
   \textstyle{ Var(\beta,\beta)=\int \big [T_{1,2,\ldots N}^{\beta}(\mathbf{x})-
    \beta\big ]^2L_{1,2,\ldots N}(\mathbf{x},\beta,\gamma) \mathrm{d} \mathbf{x}
    \geq \frac{\sum_{j=1}^N\,y_j^2/\sigma_j^2}{\sum_{j=1}^N\,y_j^2/\sigma_j^2 \sum_{k=1}^N\,1/\sigma_k^2-(\sum_{j=1}^N\,y_j/\sigma_j^2)^2 }}
\end{equation}
and:
\begin{equation}\label{eq:equation_33}
   \textstyle{ Var(\gamma,\gamma)=\int \big [T_{1,2,\ldots N}^{\gamma}(\mathbf{x})-
    \gamma\big ]^2L_{1,2,\ldots N}(\mathbf{x},\beta,\gamma) \mathrm{d} \mathbf{x}
    \geq \frac{\sum_{j=1}^N\,1/\sigma_j^2}
    {\sum_{j=1}^N\,y_j^2/\sigma_j^2 \sum_{k=1}^N\,1/\sigma_k^2-(\sum_{j=1}^N\,y_j/\sigma_j^2)^2 }}
\end{equation}
The Gaussian model is efficient because its variances are equal to the diagonal of $\mathbf{I^{-1}}$
(here and in the following the indications ${1,2,\ldots,N}$ of the used PDFs will be subtended).
The unbiased efficient estimators of $\beta$ and $\gamma$ for the model of eq.~\ref{eq:equation_25}
are given by $\mathbf{I^{-1}} U(\mathbf{x},\beta,\gamma)$ (similar to those for a single parameter):
\begin{equation}
  \textstyle{  U(\mathbf{x},\beta,\gamma)=\left (\begin{array} {c} T_1(\mathbf{x}) \\ T_2(\mathbf{x}) \end{array} \right )-\mathbf{I}\left (\begin{array} {c} \beta \\ \gamma \end{array} \right ) \ \ \ \ \  \ \ \ \ \ \
    \mathbf{I}^{-1}U(\mathbf{x},\beta,\gamma)=\mathbf{I}^{-1}\left (\begin{array} {c} T_1(\mathbf{x}) \\ T_2(\mathbf{x}) \end{array} \right )-\left (\begin{array} {c} \beta \\ \gamma \end{array} \right ) }
\end{equation}
The last equation allows the extraction of the formal expressions of the unbiased efficient
estimators for this model (and are identical to those reported everywhere
for example in ref.~\cite{particle_group}). The mean values of products of logarithmic derivatives 
are obtained by the mean values of double derivative of a single logarithmic 
function (as in eq.~\ref{eq:equation_7}), and due to eq.~\ref{eq:equation_28}  gives:
\begin{equation}\label{eq:equation_35}
\begin{aligned}
    &\int U(\mathbf{x},\beta,\gamma)\cdot U'(\mathbf{x},\beta,\gamma)\,
    L(\mathbf{x},\beta,\gamma) \mathrm{d} \mathbf{x} =\mathbf{I}  \\
    &\int \mathbf{I^{-1}} U(\mathbf{x},\beta,\gamma)\cdot U'(\mathbf{x},\beta,\gamma)
    \mathbf{I^{-1}}'\,L(\mathbf{x},\beta,\gamma) \mathrm{d} \mathbf{x} =\mathbf{I^{-1}}   \,.
\end{aligned}
\end{equation}
The second equation uses the first equation giving the Fisher
information and the symmetry $\mathbf{I^{-1}}=\mathbf{I^{-1}}'$. 
The diagonal terms are the variances of the efficient estimators for $\beta$ and
$\gamma$. It is evident the privileged position of the Gaussian PDFs in the CRF-inequality.

\subsection{The standard least-squares equations with heteroscedastic likelihood}

We defined standard least squares equations the estimators (for $\beta$ and
$\gamma$) obtained with a homoscedastic model.
It is easy to prove that the standard least squares estimators are efficient estimators for a
Gaussian model with identical $\sigma$s. The extension of eqs.~\ref{eq:equation_14} to
the case for two parameters shows immediately this property. The condition
$\sum_{i=1}^Ny^i=0$ simplifies the forms of the estimators. In the case of identical
$\sigma$s the CRF-inequality states that any other estimator has a greater variance
compared to the efficient estimators.
It is evident that the efficiency of an estimator is strictly bound to the
likelihood model of its definition (in a given model the efficient estimator
of a parameter is unique) .
Outside this model the estimators are no more efficient. We could derive the forms
of the standard least squares estimators as the efficient estimators for
the Gaussian model with identical $\sigma_j$, but the usual derivation is faster.
The estimators of $\beta$ and $\gamma$ of the standard least
squares are:
\begin{equation}
    \begin{aligned}
    &\sum_{i=1}^N\left (x_i-\beta-\gamma y_i\right )=0 \ \ \ \ \ \
    \widehat{T_s^{\beta}}(\mathbf{x},\beta,\gamma)=\frac{\sum_{i=1}^N x_i}{N}-\beta
    \ \ \ \ \\ 
    &\sum_{i=1}^N\left (x_i-\beta-\gamma y_i\right )y_i=0\ \ \ \ \ \ \
    \widehat{T_s^{\gamma}}(\mathbf{x},\beta,\gamma)=\frac{\sum_{i=1}^N x_iy_i}{\sum_{i=1}^N y_i^2}-\gamma
    \end{aligned}
\end{equation}
For an easy check  of the forms of the estimators,
we report the least squares equations that give their expressions (after inserting $\sum_i y_i=0$).
The heteroscedastic likelihood $L(\mathbf{x},\beta,\gamma)$ destroys their efficiency (optimality)
and the new variances are different from those in the homoscedastic model ($\sigma^2/N$ 
and $\sigma^2/\sum_jy_j^2$):
\begin{equation}
\begin{aligned}
    &\int \widehat{T_s^{\beta}}(\mathbf{x},\beta,\gamma)^2 L(\mathbf{x},\beta,\gamma) \mathrm{d} \mathbf{x}=
    \frac{\sum_{i=1}^N\sigma_i^2}{N^2} \\
    &\int \widehat{T_s^{\gamma}}(\mathbf{x},\beta,\gamma)^2 L(\mathbf{x},\beta,\gamma) \mathrm{d} \mathbf{x}=
    \frac{\sum_{i=1}^N\sigma_i^2 y_i^2}{(\sum_{i=1}^N y_i^2)^2}\,,
\end{aligned}
\end{equation}
and, due to the differences from the variances of the efficient estimators,
the CRF inequalities of eq.~\ref{eq:equation_32} and
eq.~\ref{eq:equation_33} impose to them to be grater.
We will extensively prove this property as a prototype of 
demonstration~\cite{econometrics} that will be used even 
for the momentum variances.

First of all, it is required the joint variance-covariance  matrix of $\widehat{T_s}^{\beta,\gamma}$
with the efficient estimators $\mathbf{I^{-1}}U(\mathbf{x},\beta,\gamma)$. This matrix is
positive semi-definite (symmetrical with the variances in the diagonal). Let us define 
with $\mathbf{C}$ this matrix, with $T$ the column vector $T=\{\widehat{T_s}^{\beta}, T_s^{\gamma}\}$
and with $<\ >$ the average on the likelihood of eq.~\ref{eq:equation_25}. The matrix $\mathbf{I^{-1}}$
is the variance-covariance matrix of the efficient estimators (eq.~\ref{eq:equation_35})
\begin{equation}
    \mathbf{C}=\left (\begin{array}{c c} <T\,T'> & <T\, U'\mathbf{I^{-1}}'> \\
                              <\mathbf{I^{-1}} U\,T'> & \ \ \ \ \mathbf{I^{-1}} \\
                       \end{array} \right )   \geq 0
\end{equation}
It is easy to show this general property of the covariance  of the
unbiased estimator $\widehat{T}_s^\beta$ with the efficient estimators. 
By definition, the mean value of an unbiased estimator is always zero 
as any of its derivative:
\begin{equation}\label{eq:equation_39}
    \frac{\partial}{\partial\beta}\int (T_s^{\beta}(\mathbf{x})-\beta) L(\mathbf{x},\beta,\gamma) \mathrm{d} \mathbf{x}=0
    =-1+\int(T_s^{\beta}(\mathbf{x})-\beta) \frac{\partial}{\partial\beta}L(\mathbf{x},\beta,\gamma) \mathrm{d} \mathbf{x}\,.
\end{equation}
The derivative of the likelihood is the first component of $U(\mathbf{x},\beta,\gamma)$
of eq.~\ref{eq:equation_27}, and this correlation becomes equal to one. The derivative in $\gamma$ of
the mean value of $\widehat{T}_s^\beta$ has no one and the sought correlation is zero.
Similarly for the other estimator $\widehat{T}_s^\gamma$. Hence the matrix  $ < T\, U' >$ is:
\begin{equation}
    <T\, U'>=\mathbb{I}_2=\left (\begin{array}{c c} 1 & 0 \\
                                                    0 & 1 \\
                                \end{array} \right )      = <U\, T'>
                                \ \  \Rightarrow \ \ \mathbf{C}=\left (\begin{array}{c c} <T\,T'> & \mathbf{I^{-1}} \\
                              \mathbf{I^{-1}} & \ \ \ \ \mathbf{I^{-1}} \\
                       \end{array} \right )  \geq 0
\end{equation}
The $\mathbf{I^{-1}}$ is a symmetric matrix. The positivity of $\mathbf{C}$ is
conserved even with transformation $\mathbf{A}'\,\mathbf{C}\,\mathbf{A}$
with $\mathbf{A}$ any real matrix of four lines and two columns (few details 
about the positive definite matrices are reported in the appendix):
\begin{equation}\label{eq:equation_41}
    \left ( \mathbb{I}_2\ \ -\mathbb{I}_2\right ) \mathbf{C}
    \left (\begin{array} {c } \,\, \mathbb{I}_2  \\
                           - \mathbb{I}_2
    \end{array} \right) = <T\, T'> - \mathbf{I^{-1}} \geq 0 \ \ \ \ \ \
    \Rightarrow \ \ \ \ \ \  \big\langle T\, T'\rangle \  > \ \mathbf{I^{-1}}
\end{equation}
The last of the eq.~\ref{eq:equation_41} is eq.~\ref{eq:equation_31} as a matrix relation
the equality is eliminated being evidently impossible.
For the definition of positive definite matrices it is straightforward to extract the relations
among the variances. The eqs.~\ref{eq:equation_31} to~\ref{eq:equation_41} will be
extended the estimators of the momentum reconstructions.

Even if it was demonstrated
with a heteroscedastic Gaussian model,  eqs.~\ref{eq:equation_41}   remain always
greater also for heteroscedastic irregular models with equivalent
set of $\{\sigma_j\}$. The efficient estimators of
heteroscedastic models continue to be efficient even for homoscedastic models,
converging to them when all the $\sigma_j$ become identical.

\begin{figure} [h!]
\begin{center}
\includegraphics[scale=0.75]{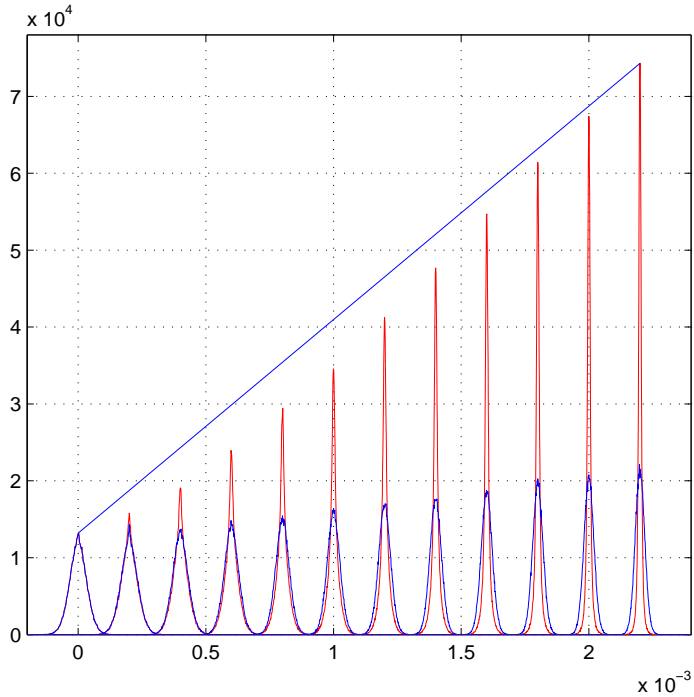}
\caption{\em The Gaussian model. A faster increase in the maximums compared to that of fig.1.
The blue straight lines connect the maximums of the first distributions to the last red distribution.
The color cade is that of fig.1. }
\label{fig:figure_3}
\end{center}
\end{figure}

All this set of demonstrations for heteroscedastic Gaussian model is addressed
to each given  chain of PDFs with a defined order of $\sigma_j$. Their
variances of eq.~\ref{eq:equation_32} and
eq.~\ref{eq:equation_33} are strongly modified by the differences
among the $\{\sigma_j\}$ without any evident $N$ limitation.
Instead, the variances of the standard least
squares are weakly modified and they save
explicit $N$ limitations. For example the variances of the
$\beta$ parameter has averages $\approx \sigma_m^2/N$.
Figure~\ref{fig:figure_1} illustrates this limits for the $\gamma$
estimator (the blue distributions).
Due to the heavy modifications of the variances of the chains
of PDFs,  other PDFs become essential.
For example those that select  the chains among the
possible set of chains. These PDFs are introduced by the
properties of the measuring devices.
As we said above, a binomial PDF suffices in the disordered tracker
detector for the models of refs.\cite{landi07,landi06}, just
a mean disorder of half strip ($\approx\, 30\, \mu m$) is enough
for this.
%
The use of standard least squares method in
heteroscedastic models has enormous simplifications,
because it does not need the knowledge of the
$\sigma_j$ for each observation.
Unfortunately, it amounts to a large loss of resolution.
Our models of ref.~\cite{landi07} illustrate in evident way
the amplitude of this loss. In any case the lucky model of
ref.~\cite{landi07} can be an economic way to introduce effective
$\sigma_j$.

For our pleasure, in fig.~\ref{fig:figure_3}, we invented a faster
growth than that of the Gaussian model of ref.~\cite{landi07}. The
quality of the detectors increases with the number N of detecting
layers, each inserted layer
is better than the previous one. The bleu straight line shows an evident
deviation from a linear growth with an almost parabolic growth. The
blue distributions are practically insensible to these changes. With
cleaver selection of the hit quality many types of growth can be implemented.

\subsection{Deviations from optimality}

Equations~\ref{eq:equation_41} proved the non-optimality of the standard least-squares
in the case of it use with a different (heteroscedastic) likelihood. The standard least
squares gives the efficient estimators for a homoscedastic Gaussian model, the 
transportation outside its condition of efficiency (optimality) destroys its optimality
and produces an increase of variances with a degradation of the resolution as illustrated
by the simulations of ref.~\cite{landi07,landi06,landi05}. Even our 
heteroscedastic model suffers of a similar non-optimality loss outside its range of validity.
We underlined that the orders and the values of the sets of $\{\sigma_j\}$ define the model,
each deviation from the order and values implies an increase of the variances compared 
to the optimality condition. We used this property to prove the essential connection of 
the appropriate $\sigma_j$ with the hit $j$. If the ordered set of $\{\sigma_j\}$ 
is randomized, destroying the correlation with the set of the hits, the 
distributions of the estimators change drastically becoming worse than the 
standard least squares. Instead, the differences of the heteroscedastic 
efficient estimator minus a suboptimal estimator with few eliminated PDFs rapidly 
converge as the eliminated PDFs are reduced. This rapid convergence is not 
observed in the case of the standard least-squares.

\section{Inequality for Momentum estimators}

As stated above the CRF-inequality can be extended to case of the momentum
reconstruction. It will be studied the case of a high-momentum  
charged-particle moving in a tracker with a homogeneous
magnetic field and the path of the particle is  
orthogonal to the field direction. Here,
the circular path can be approximated with a parabola. As for the
straight tracks we will compare the estimators of standard least
squares with those of a heteroscedastic model. The estimators of
the standard least squares are efficient (optimum) in the homoscedastic
Gaussian model.

\subsection{The momentum estimators in homoscedastic Gaussian model}

We use this approach to define our expressions for the estimators. These
forms will be used in heteroscedastic models. Our equation will be expressed
in compact way with matrices. The function $\mathfrak{L}(\mathbf{x},\beta,\gamma,\eta)$
is the likelihood for this case, $\beta$ and $\gamma$ are just defined,
$\eta$ is the curvature of the track, proportional to $1/p$ with $p$
the track momentum:
\begin{equation}
\begin{aligned}
    &\mathfrak{L}(\mathbf{x},\beta,\gamma,\eta)=f_1(x_1,\beta,\gamma,\eta)f_2(x_2,\beta,\gamma,\eta)
    \ldots\,f_N(x_N,\beta,\gamma,\eta) \\
    &f_j(x,\beta,\gamma,\eta)=\frac{\exp[-(x-\beta-\gamma y_j -\eta y_j^2)^2/2\sigma^2]}{\sqrt{2\pi}\sigma}
\end{aligned}
\end{equation}
The index $j$ orders the PDFs $f_j$ as the tracker layers.
The equivalent of eq.~\ref{eq:equation_27} is:
\begin{equation}
    V(\mathbf{x},\beta,\gamma,\eta)=\left (\begin{array} {l}
    \displaystyle{\frac{\partial\ln \mathfrak{L}(\mathbf{x},\beta,\gamma,\eta)}{\partial\beta}}\\
    \displaystyle{\frac{\partial\ln \mathfrak{L}(\mathbf{x},\beta,\gamma,\eta)}{\partial\gamma}}\\
    \displaystyle{\frac{\partial\ln \mathfrak{L}(\mathbf{x},\beta,\gamma,\eta)}{\partial\eta}}
     \end{array} \right )
\end{equation}
The unbiased estimators for $\beta,\gamma,\eta$ are:
\begin{equation}
    \textstyle{  V(\mathbf{x},\beta,\gamma,\eta)=\left (\begin{array} {c} T_a(\mathbf{x}) \\ T_b(\mathbf{x})
    \\ T_c(\mathbf{x})\end{array} \right )-\mathbf{R}\left (\begin{array} {c} \beta \\ \gamma \\ \eta \end{array} \right ) \ \ \ \ \  \ \ \ \ \ \
    \mathbf{R}^{-1}V(\mathbf{x},\beta,\gamma,\eta)=\mathbf{R}^{-1}\left (\begin{array} {c} T_a(\mathbf{x}) \\ T_b(\mathbf{x})
    \\ T_c(\mathbf{x})\end{array} \right )-\left (\begin{array} {c} \beta \\ \gamma \\ \eta\end{array} \right ) }
\end{equation}
The vector $\mathbf{R}^{-1}V(\mathbf{x},\beta,\gamma,\eta)$ contains the efficient estimators of this model,
the matrix $\mathbf{R}$ is its the Fisher information (with the condition $\sum_j y_j=0$ and $j=\scriptstyle{\{1,2,\ldots,N\}}$):
\begin{equation}
    \mathbf{R}=\left (\begin{array} {l l l}
                     \textstyle{N } &  \textstyle{0} & \textstyle{\sum_{j} y_j^2}\\
                      \textstyle{0} &  \sum_{j} y_j^2 & \sum_{j} y_j^3\\
                      \sum_{j} y_j^2 & \sum_{j} y_j^3 & \sum_{j} y_j^4 \end{array} \right ) \frac{1}{\sigma^2}
\end{equation}
For eq.~\ref{eq:equation_35} extended to this case, it is:
\begin{equation}
\begin{aligned}
    &\int V(\mathbf{x},\beta,\gamma,\eta)\cdot V'(\mathbf{x},\beta,\gamma,\eta)\,
    \mathfrak{L}(\mathbf{x},\beta,\gamma,\eta) \mathrm{d} \mathbf{x} =\mathbf{R}  \\
    &\int \mathbf{R^{-1}} V(\mathbf{x},\beta,\gamma,\eta)\cdot V'(\mathbf{x},\beta,\gamma,\eta)
    \mathbf{R^{-1}}'\,\mathfrak{L}(\mathbf{x},\beta,\gamma,\eta) \mathrm{d} \mathbf{x} =\mathbf{R^{-1}}   \,.
\end{aligned}
\end{equation}
The matrices $\mathbf{R}$ and $\mathbf{R^{-1}}$ are symmetric and positive semi-defined, and
the second line of the last equation is the variance-covariance matrix of the efficient estimators.
The variance-covariance matrix of a generic vector $T$ of unbiased estimators
of $\beta,\gamma,\eta$ with the efficient estimators is:
\begin{equation}\label{eq:equation_47}
    \mathbf{K}=\left (\begin{array}{c c} <T\,T'> & <T\, V'\mathbf{R^{-1}}'> \\
                              <\mathbf{R^{-1}} V\,T'> & \ \ \ \ \mathbf{R^{-1}} \\
                       \end{array} \right )   \geq 0
\end{equation}
As defined, the symbol $<>$ means an integral with the likelihood $\mathfrak{L}$, the cross
correlations $<V\,T'>$ follow the rules of eq.~\ref{eq:equation_39} giving now the $3x3$ identity
matrix $\mathbb{I}_3$ reducing the off-diagonal elements of $\mathbf{K}$ to $\mathbf{R}^{-1}$.
With a transformation of the type $\mathbf{A'}\,\mathbf{K}\,\mathbf{A}$ with a $6x3$ real matrix
$\mathbf{A}$ the fundamental inequality is obtained:
\begin{equation}\label{eq:equation_48}
    \left ( \mathbb{I}_3\ \ -\mathbb{I}_3\right )\left (\begin{array}{c c} <T\,T'> & \mathbf{R^{-1}}' \\
                              \mathbf{R^{-1}} & \ \ \ \ \mathbf{R^{-1}} \\
                       \end{array} \right )\left (\begin{array} {c } \,\, \mathbb{I}_3  \\
                           - \mathbb{I}_3
    \end{array} \right) = \langle T\, T'\rangle - \mathbf{R^{-1}} \geq 0 \ \ \ \ \ \
\end{equation}
The equality is obtained only for the efficient estimators of this model. Therefore,
any other set of estimators have a greater variance in this model. Any heteroscedastic
model has greater variance than the efficient model. The reason is connected to the
modification that the likelihood $\mathfrak{L}$ introduces in the variances. These
modifications always increase the variances. In this model, the parameter $\sigma$
disappears in the expressions for the estimators. The form $\mathbf{R}$ of the
Fisher information depends from N, this limits the growth with the numbers of
detecting layers as illustrated in ref.~\cite{landi06}.

\subsection{The momentum estimators in heteroscedastic Gaussian model}

In its essence this model can be obtained from the corresponding homoscedastic model with
a set of formal differences.
\begin{equation}
\begin{aligned}
    &{L}(\mathbf{x},\beta,\gamma,\eta)=f_1(x_1,\beta,\gamma,\eta)f_2(x_2,\beta,\gamma,\eta)
    \ldots\,f_N(x_N,\beta,\gamma,\eta) \\
    &f_j(x,\beta,\gamma,\eta)=\frac{\exp[-(x-\beta-\gamma y_j -\eta y_j^2)^2/2\sigma_j^2]}{\sqrt{2\pi}\sigma_j}
\end{aligned}
\end{equation}
The index $j$ orders the PDFs as the detecting layers and selects a different $\sigma_j$ for this observation.
The other differences are in the definitions of the vector $T_1,T_2,T_3$ that are different from the previous
$T_a,T_b,T_c$ and the Fisher information $\mathbf{I}$.
\begin{equation}
    \textstyle{  U(\mathbf{x},\beta,\gamma,\eta)=\left (\begin{array} {c} T_1(\mathbf{x}) \\ T_2(\mathbf{x})
    \\ T_3(\mathbf{x})\end{array} \right )-\mathbf{I}\left (\begin{array} {c} \beta \\ \gamma \\ \eta \end{array} \right ) \ \ \ \ \  \ \ \ \ \ \
    \left (\begin{array} {c} T_1(\mathbf{x}) \\ T_2(\mathbf{x})
    \\ T_3(\mathbf{x})\end{array} \right )=\left (\begin{array} {c} \sum_j\frac{ x_j}{\sigma_j^2} \\ \sum_j\frac{x_j\,y_j}{\sigma_j^2}
    \\ \sum_j\frac{x_j\,y_j^2}{\sigma_j^2} \end{array} \right ) }
\end{equation}
The Fisher information becomes:
\begin{equation}
    \mathbf{I}=\left (\begin{array} {l l l}
                     \sum_j {1/\sigma_j^2}  &  \sum_j{y_j/\sigma_j^2} & \sum_{j}{ y_j^2/\sigma_j^2}\\
                      \sum_j{y_j/\sigma_j^2} &  \sum_{j}{y_j^2/\sigma_j^2} & \sum_{j}{y_j^3/\sigma_j^2}\\
                      \sum_{j}y_j^2/\sigma_j^2 & \sum_{j} y_j^3/\sigma_j^2 & \sum_{j} y_j^4/\sigma_j^2 \end{array} \right )
\end{equation}
The efficient unbiased estimators for the likelihood ${L}(\mathbf{x},\beta,\gamma,\eta)$
are contained in the vector $\mathbf{I^{-1}}U(\mathbf{x},\beta,\gamma,\eta)$. 
Even in this case the Fisher information
has no obvious N-dependence. The variance-covariance matrix for any vector $T$ of unbiased
estimators can be transformed as previously described giving the inequality:
\begin{equation}\label{eq:equation_52}
    \left ( \mathbb{I}_3\ \ -\mathbb{I}_3\right )\left (\begin{array}{c c} <T\,T'> & \mathbf{I^{-1}}' \\
                              \mathbf{I^{-1}} & \ \ \ \ \mathbf{I^{-1}} \\
                       \end{array} \right )\left (\begin{array} {c } \,\, \mathbb{I}_3  \\
                           - \mathbb{I}_3
    \end{array} \right) = \langle T\, T'\rangle - \mathbf{I^{-1}} \geq 0 \ \ \ \ \ \
\end{equation}
Here, the symbol $<\,>$ indicates an integral with the likelihood of this model
${L}(\mathbf{x},\beta,\gamma,\eta)$. The equality to zero is obtained only when
the vector $T$ is $\mathbf{I^{-1}}U(\mathbf{x},\beta,\gamma,\eta)$. For any other unbiased
estimator, that difference is greater than zero (in the sense of the positive definite
matrices).

\section{Conclusions}

The Cramer-Rao-Frechet inequality is applied to a set of heteroscedastic Gaussian model to
prove the absence of any limitation introduced by this fundamental inequality. Instead,
the inequality shows that, in heteroscedastic models, the estimators of the standard 
least-squares have always greater variances compared to that given by
the weighted least-squares. The variances of the standard least-squares
conserve approximate $1/N$ form that imply limitations in the growth of the
distributions with the number $N$ of detecting layers. No similar limitations are
present in the weighted least squares that are dominated by the differences of the
weights $1/\sigma_j^2$. For identical $\sigma_j$ the estimators converge to those of
the standard least-squares.
The inequality is calculated also for the momentum estimators. Again the
appropriate heteroscedastic estimators have a lower variances than those of the
standard least squares. Given the loss of resolution implied by greater variances it
must be avoid the use of standard least squares with heteroscedastic data. The lucky model
may be helpful in silicon detectors.

\section{Appendix}

Let us add few details about variance-covariance matrix or about the cross-covariance matrix. For
their definitions these matrices are positive semi-defined (they are symmetric with the 
positive variances in the principal diagonal). If the matrix $\mathbf{C}$ is a positive semi-definite 
matrix it must be a symmetric $mxm$ matric $\mathbf{C'}=\mathbf{C}$, and for any real 
vector $z$ it must be $z'\mathbf{C}z\geq 0$. 
From these definitions it follows that if $\mathbf{A}$ is a real matrix of $mxn$ with $n<m$
the matrix $\mathbf{K}$ given by $\mathbf{A'}\,\mathbf{C}\,\mathbf{A}$ is again a 
$nxn$ positive semi-definite matrix. It is easy to show that the diagonal elements of $\mathbf{K}$
are given by $z'\mathbf{C}z$ where each  $z$ is a column of $\mathbf{A}$ and the off-diagonal 
terms are symmetric. A positive semi-definite matrix implies an infinite number 
 of inequalities ($\infty^m$),
our interest is to find the minimal one with useful dimensions. 
The eqs.~\ref{eq:equation_41}~\ref{eq:equation_48}~\ref{eq:equation_52} are addressed to this.
It is easy to show that the identity matrices $\mathbb{I}_2$ and $\mathbb{I}_3$ select the
useful blocks of the matrices. AS an example we can elaborate the $\mathbf{C}$ matrix ($4x4$).  
In general one can use a $4x2$ matrix of the form
$(\mathbb{I}_2,a\mathbb{I}_2)$ with $a$ any real constant and search the minimum in $a$:
\begin{equation}
\begin{aligned}
    &\left ( \mathbb{I}_2\ \ a\mathbb{I}_2\right ) \mathbf{C}
    \left (\begin{array} {c } \,\, \mathbb{I}_2  \\
                           a \mathbb{I}_2
    \end{array} \right) = <T\, T'> +(2\,a +a^2) \mathbf{I^{-1}} \geq 0   \\
    & \frac{\partial}{\partial a}\left(<T\, T'> +(2\,a +a^2) \mathbf{I^{-1}} \right )\ \ \ \Rightarrow \ \ (2+2 a)=0 \ \ \ \ \
    \Rightarrow \ \ \ \ \ \ \mathbf{B}= < T\, T'> - \mathbf{I^{-1}} \geq 0
\end{aligned}
\end{equation}
The solution $a=-1$ gives  the last inequality.
The inequality for each couple of variances is obtained with  $z'\mathbf{B}\,z$ where
$z$ is $(1,0)$ for the first two and $(0,1)$ for the second two.


\end{document}